\documentclass[12pt]{article}
\usepackage[centertags]{amsmath}
\usepackage{amsfonts}
\usepackage{bm}
\begin{document}
\title{\text \bf{Real-Space  Renormalization Group for Quantum Gravity I:\\ 
Significance of terms linear and quadratic in curvature}}
\author{H.S.Sharatchandra\thanks{E-mail:
{sharat@cpres.org}} \\[2mm]
{\em Centre for Promotion of Research,} \\
{\em 7, Shaktinagar Main Road, Porur, Chennai 600116, India}}
\date{}
\maketitle
\begin{abstract}
Real-Space  renormalization group techniques are developed for tackling large curvature fluctuations in quantum gravity. Within cells of invariant volume $a^4$, only certain types of fluctuations are allowed. Normal coordinates are used to 
avoid redundancy of the degrees of freedom. The relevant integration measure is read off from the metric on metrics. All  fluctuations in a group of cells are averaged over to get an effective action for the larger cell. In this paper the simplest type of fluctuations are kept. The measure is simply an integration over independent components of the curvature tensor at the center of each cell. Terms of higher order in $a$ are required  for convergence in case of Einstein-Hilbert action.
With only next order (in $a$) contribution to the  action, there is no renormalization of Newton's or cosmological constants. 
The `massless Gaussian surface' in the renormalization group space is given by actions that have  linear and quadratic terms in curvature and determines the evolution of coupling constants away from it. Our techniques allow for systematic improvements.
\end{abstract}

\smallskip

\smallskip


\section{Introduction}\label{i}

Einstein's general theory of relativity has had phenomenal  experimental successes as a classical theory. But quantizing it  turns out to be a tough nut to crack.
Gravity with the Einstein-Hilbert action is non-renormalizable. The loop corrections are divergent in perturbation theory,  the degree of divergence increasing with the number of loops. This means the gravitons of large frequencies have increasing and uncontrolled contributions. Using a cut-off  and adding counterterms only results in newer and newer counterterms with larger and larger coupling constants. We have an infinite number of coupling constants to be specified. The unitarity of the resulting theory also becomes an issue. All this means we have to go beyond perturbation theory, however small Newton's constant is. We have to figure out the effects of virtual gravitons of large frequencies in a non-perturbative way. These correspond to large fluctuations of (invariants of) curvature tensor and its derivatives.

In this Paper real-space  renormalization group techniques \cite{k},\cite{w},\cite{kw},\cite{dg},\cite{m}   are developed for tackling large curvature fluctuations in quantum gravity.
It is useful to draw a close parallel to our approach. Consider a surface with microscopic pits and bumps.  For a probe which has a much weaker resolution, for example, the naked eye, the surface looks smooth. Imagine a statistical problem involving averaging over all such  surfaces with a Boltzmann weight.  Microscopic fluctuations  would be relevant  for the physical properties of the surface such as color, surface tension etc.. We would like to average over the effects of pits and bumps to get an equivalent problem which involves only summing over `smooth' surfaces with an
 effective Boltzmann weight.

In case of quantum gravity there are curvature fluctuations at scales far beyond the Planck scale, while our probe could be at the scale of a meter.  When perturbation theory gives uncontrolled contributions at high frequencies,
we need to apply  renormalization group techniques. Once we get an equivalent action, there are no fluctuations smaller than a  macroscopic  length. There is an effective cutoff, and perturbation theory can be safely applied.

Our approach is as follows. We keep only certain types of fluctuations within cells of invariant volume $a^4$. This is a way of imposing a cut-off of scale $1/a$. Riemann normal coordinates \cite{g},\cite{n}  are used to 
avoid redundency of the degrees of freedom. This way we are working directly with physical lengths and physically relevant  variables, in addition to `fixing a gauge'.  The relevant integration measure for our fluctuations  is calculated  from the metric on metrics. This can be done for each cell separately. Next we consider a group of cells and define its degrees of freedom in terms of those of the individual cells. All fluctuations of these  cells are averaged over to get an effective action  for the larger cell.

In recent years there has been extensive work \cite{kpsk},\cite{p},\cite{i},  on `exact' or `functional'  renormalization group applied also to quantum gravity. The idea is to integrate over higher frequency modes in various truncation
 and approximation schemes to get an effective action for their effects on the low frequency modes. From this we can obtain  the evolution of the coupling constants, to understand the infrared behavior or look for the continuum limit. There are attempts to seek ``asymptotic safety". We are in effect doing the same, but we are tackling large curvature fluctuations more directly.

\section{Restricting the fluctuations in a cell}\label{f}
We are interested in tackling the partition function
\begin{equation}\label{z}
Z=\int D g\ e^{iS[g]}\\
\end{equation}
with the Einstein-Hilbert action:
\begin{equation}\label{eh}
S = \frac{1}{2\kappa}\int d^{D}x \sqrt{-g}(R-2\Lambda)
\end{equation}
where $\kappa=8\pi G$, $G$ is the Newton's constant and $\Lambda$, the cosmological constant. We have to integrate over all metrics $g_{\mu \nu}$ over space-time.
Here $\mu, \nu= 0,1,2,3$ and we choose the signature $(-+++)$ for the metric.

Consider a division of space-time into `cells', each of an {\bf invariant volume} $a^4$. The length scale $a$ is inverse of the cutoff. Within each cell we freeze fluctuations except of a specific type.  This is best formulated using normal coordinates \cite{g} around the `center' of the cell. This has an additional advantage of `fixing the gauge' (except for a rigid Lorentz transformation) within each cell.
The metric $g_{\alpha \beta}(x)$ in each cell has the following Taylor expansion. 
\begin{equation}\label{nc}
g_{\alpha \beta}(x)=\eta_{\alpha \beta}+\frac{1}{3}R_{\alpha \gamma \delta\beta}x^\gamma x^\delta+\frac{1}{ 3}R_{\alpha \gamma \delta\beta,\mu}x^\gamma x^\delta x^\mu+\frac{1}{5}(6 R_{\alpha \gamma \delta\beta,\lambda,\mu}+ \frac{16}{3} R_{\alpha \gamma \delta \rho}R^{\rho}_{\lambda \mu \beta})x^\gamma x^\delta x^\lambda x^\mu +\cdots.\\
\end{equation}
Here  the  curvature tensor at the center of the cell is denoted by the 4-index numbers  $R_{\alpha \gamma \delta\beta}$,
and the values of  its   covariant derivatives there by the numbers $R_{\alpha \gamma \delta\beta,\mu, \nu}$ etc. 

In this paper in each cell $c$ we allow only the metrics  
\begin{equation}\label{mc}
g_{\alpha \beta}(x)=\eta_{\alpha \beta}+\frac{1}{3}R_{\alpha \gamma \delta\beta}x^\gamma x^\delta+ `small'.
\end{equation}
Note that we are allowing for arbitrarily large  curvature (invariants) within the cell.

The algebraic curvature tensor   $R_{\alpha \gamma \delta\beta}$  have the symmetry properties
\begin{eqnarray}\label{rt1}
&&R_{\alpha \gamma \delta\beta}=R_{\delta\beta\alpha \gamma },\nonumber\\
&&R_{\alpha \gamma \delta\beta}=-R_{\gamma\alpha  \delta\beta}=-R_{\alpha \gamma \beta\delta}
\end{eqnarray}
and also  the Bianchi identity restriction 
\begin{eqnarray}\label{rt2}
R_{\alpha \gamma \delta\beta}+R_{\delta\alpha \gamma \beta}+R_{ \gamma \delta\alpha\beta}=0.
\end{eqnarray}
In four space-time dimensions the only independent restriction from this identity is
\begin{eqnarray}\label{rt3}
R_{1234}+R_{3124}+R_{2314}=0
\end{eqnarray}
resulting in $20$ independent components.

If we ignore all derivative terms in the Taylor expansion  (\ref{nc}), it is equivalent to having all covariant derivatives vanish within the cell. We then have a neighborhood of a maximally symmetric space within the cell. Then the algebraic curvature tensor  $R_{\alpha \gamma \delta\beta}$ has the form $R (g_{\alpha \delta}g_{ \gamma \beta}- g_{\alpha \beta }g_{ \gamma \delta} )$, where $R$ is the curvature scalar.  Then we are only allowing fluctuations of the metric labeled by only the curvature  scalar  within the cell. This is a little too restrictive for us. Therefore we are allowing for deviations from this restriction, but restricting them to be 'small' within each cell. Our techniques allow for these restrictions to be modified.

\section{Measure from metric on metrics}\label{mm}
We now compute integration measure for the metrics within each cell using the metric on metrics. For a infinitesimal metric $\delta g_{\mu\nu}(x)$ the diffeomorphism invariant metric is
\begin{eqnarray}
<\delta g, \delta g>=\int d^4 x\sqrt{-g}\, \delta g_{\mu\nu}g^{\mu\rho} g^{\nu\sigma}\, \delta g_{\rho\sigma}.
\end{eqnarray}
For ease of calculation we want to be able to use Lorentz invariance  at each step. As the Lorentz group is non-compact, we cannot choose cells of finite invariant volumes which are invariant under the group. Therefore we do calculations for the Euclidean case and translate the results for our Minkowski case. We  choose our cell as an 4-ball of radius $b$ as this is invariant under  an  $SO(4)$ rotation about its center. We have for volume of our `cell' $a^4=\pi^2 b^4/2$.
For our metrics within each cell, to the lowest order in $a$,
\begin{eqnarray}
<\delta g, \delta g>=\frac{1}{9}\int_c d^4 x \, \delta R_{iklj} x^k x^l  \delta R_{imnj} x^m x^n,
\end{eqnarray}
since  $ g_{ij}=\delta_{ij}$ at the center of the cell. Now $i,j,...$ take values $1,2,3,4$.
We have,
\begin{eqnarray}
\int_c d^4 x \, x^k x^l  x^m x^n = v_4a^8(\delta_{kl}\delta_{mn}+\delta_{km}\delta_{ln}+\delta_{kn}\delta_{lm}),
\end{eqnarray}
where
\begin{eqnarray}
v_4a^8=\frac{1}{24}\int_c d^4 x \, (x^2)^2 = \frac{\pi^2}{24}\int_0^{b^2} dr^2 (r^2)^3=\frac{\pi^2}{96}b^8.
\end{eqnarray}
We get
\begin{eqnarray}
<\delta g, \delta g>_c=\frac{3v_4a^8}{2}\delta R_{ijkl}\,\delta R_{ijkl},
\end{eqnarray}
by using the identity 
\begin{eqnarray}\label{rt4}
 R_{ijkl}\, R_{ikjl}=\frac{1}{2}R_{ijkl}\, R_{ijkl},
\end{eqnarray}
which follows from the  Bianchi identity and which is also valid for $\delta R_{ijkl}$.
Therefore the measure is simply an integration over each independent component of the curvature tensor over the entire range $(-\infty,\infty)$ subject only to one Bianchi constraint. Thus it is a $20$ dimensional integration. 
Therefore we get the for the measure over our metrics (\ref{mc}) in the cell $c$, 
\begin{eqnarray}\label{20} 
\int_c \, Dg=\int \, d^{20}R.
\end{eqnarray}
It is useful to decompose the curvature tensor into the Weyl tensor $C_{ijkl}$, the Ricci scalar $R$, and the traceless part 
of the Ricci  tensor,
\begin{eqnarray}
\hat R_{\mu\nu}=R_{\mu\nu}-{1\over 4}g_{\mu\nu}R.
 \end{eqnarray}
We have,
\begin{eqnarray}\label{rt5}
R_{ijkl} R_{ijkl}=C_{ijkl} C_{ijkl}+2\hat R_{ij} \hat R_{ij}+\frac{1}{6}R^2.
\end{eqnarray}
Therefore the integration measure for each cell is  
\begin{eqnarray}\label{mw}
\int_c D g = \int  dR \, d^9\hat R_{ij} \, d^{11} C_{klmn} \, \delta(C_{1234}+C_{3124}+C_{2314}).
\end{eqnarray}
We are integrating over all independent components of $\hat R_{ij}, C_{klmn}$. The Bianchi identity constraint for the Weyl tensor is explicitly written with a Dirac $\delta$ function.

\section{Coarse-graining}\label{cg}
The measure Eq.(\ref{mw})  raises a serious  issue. With our choice of fluctuations in the cell, the lowest order (in $a$) contribution to the Einstein-Hilbert action is
\begin{eqnarray}
\frac{1}{2\kappa}\int_{c}d^{4}x \sqrt{-g}(R-2\Lambda)|_0=\frac{a^4}{2\kappa}(R-2\Lambda)
\end{eqnarray}
Thus there is no convergence for integration over our algebraic  Weyl and Ricci tensors,  $C_{klmn},\hat R_{ij}$. Also in absence of matter or external sources, the integration over the curvature scalar $R$ gives a vanishing partition function!

We might interpret this to mean that the Einstein theory cannot be the microscopic theory or that it does not make sense without matter. We might also interpret that we need a cutoff for the range of values the components of the curvature tensor (in addition to the truncation of fluctuations within a cell) and use renomalization group to remove this cutoff too. However we regard it to mean that we need to keep terms of higher order in $a$ in the action.

The contribution of next order terms in $a$ to the action are calculated in the Appendix. Converting the result to the Lorentz case,
\begin{eqnarray}
\int_c d^4 x\sqrt{-g}\, S|_2=a^6\Big{(}\kappa_1 R^2+\kappa_2\hat R_{\mu\nu}^2+\kappa_3 C_{\mu\nu\rho\sigma}C^{\mu\nu\rho\sigma}\Big{)}. 
\end{eqnarray}
As a consequence of the quadratic terms in the curvature tensor, now the integrals exist. These terms are as if there are counterterms that are quadratic in $a$,  (the cut-off being $a^{-1}$)  and quartic in derivatives but with coefficients that are fixed  and cannot be adjusted at will. Also we have obtained them from the  Einstein-Hilbert action and we have not tampered with the unitarity.

We now apply renormalization group procedure to our problem. We consider a larger cell $C$ made of $N$ of these smaller cells $c_n, n=1,2,..,N$. We average over all fluctuations of the smaller cells consistent with a ``smooth"  metric of the larger cell. Specifically, we integrate over all $R_{\mu\nu\rho\sigma}(c)$ of the smaller cells giving a mean curvature  of the larger cell:
\begin{eqnarray}\label{m}
R_{\mu\nu\rho\sigma}(C)=\frac{1}{N}\sum_n R_{\mu\nu\rho\sigma}(c_n).
\end{eqnarray}

Actually it is not correct to add up the components of the curvature tensor at different points, especially because we have used normal coordinates separately within different cells. At the least, we have to use normal coordinates over the entire cell $C$ centered around its `center'. In this paper we presume that this is done. Another way is to use curvature invariants, but this is much more complicated. Implementing Eq. (\ref{m}) by a Dirac $\delta$ function the partition function for the large cell $C$ is 
\begin{eqnarray}
Z(C)=\int d^{20}R(C) e^{iS(C)},
\end{eqnarray}
where
\begin{eqnarray}
&&e^{iS(C)}=\prod_n\int d^{20}R(c_n)\, \delta\Big{(}R_{\mu\nu\rho\sigma}(C)
-\frac{1}{N}\sum_n R_{\mu\nu\rho\sigma}(c_n)\Big{)}\times\\ \nonumber
&&exp \Big{(}i\frac{1}{2\kappa}\sum_n \big{(}(a^4(R(c_n)-2\Lambda)
+a^6(\kappa_1 R^2+\kappa_2\hat R_{\mu\nu}^2+\kappa_3 C_{\mu\nu\rho\sigma}C^{\mu\nu\rho\sigma}(c_n))\big{)}\Big{)}.
\end{eqnarray}
The quadratic integrals can all be evaluated explicitly. $S(C)$ turns out to have exactly the same form as  $S(c)$, only multiplied by $N$ as required by the larger volume of the cell $C$. Even the Bianchi identity for the cells $c_n$ gives Bianchi identity for the cell $C$ because the identity is linear. All this can be illustrated  by the simple integral,
\begin{eqnarray}
\prod_n\int dx_n e^{i\sum_n(\frac{c}{2} x_n^2+bx_n)}\delta(X-\frac{1}{N}\sum x_n).
\end{eqnarray}
Using the integral representation for the Dirac $\delta$ function this is 
\begin{eqnarray}
e^{ibNX}\int \frac{dt}{2\pi} e^{-itX}\Big{(} \int dx_n e^{i ( \frac{c}{2} x_n^2+\frac{1}{N}tx_n)}\Big{)}^N
=e^{ibNX}\int dt e^{-itX}e^{-i\frac{1}{2cN}t^2}=e^{iN(\frac{c}{2}X^2+bX)}.
\end{eqnarray}
Thus we get the same form in the mean $X$ except that it is scaled by $N$.
\begin{eqnarray}\label{mC}
Z(C)=\int d^{20}R(C) e^{S(C)}
\end{eqnarray}
By using our derivation (\ref{20}) of the measure for metrics in the cell in the reverse direction,  we can interpret  Eq.(\ref{mC}) to mean that the larger cell also has only our restricted metrics (\ref{mc}), and we are integrating over them.
This way we have obtained an equivalent partition function which has a metric which is ``smooth"  over a larger cell $C$.

Thus in our approximation, the effective action over the large cell $C$ is exactly the same as for each small cell, except of coarse
the volume of the cell is now $Na^4$. This means there is no renormalization of Newton's constant or the cosmological constant!
This is of course due to keeping only the quadratic terms in the contribution to the action.

\section{Discussion}\label{c}
In this paper we have initiated direct calculation of the effects of large curvature fluctuations in quantum gravity. We have implemented a  cutoff by restricting the metric fluctuations in a cell of invariant volume $a^4$, still allowing arbitrarily large curvatures. By working with normal coordinates within each cell, we are well grounded with physically relevant fluctuations. Our truncations and approximations can be systematically improved. 

In quantum gravity we integrate over the metrics, but the physical fluctuations are in curvature. Our techniques make use of this effectively, converting the coarse-graining procedure of renormalization group into an almost ultra-local calculation.

We were forced to keep next order contribution in $a$ to the action in order to get convergence for Einstein-Hilbert action. They are quadratic in curvature, reminiscent of $R^2$ gravity, but were not put by hand. In spite of being of a higher order in $a$, they gave a finite and meaningful contribution. In deed  they led to same action for the larger cell, scaled only by the larger volume. As a result there is 
non-renormalization of both Newton's constant and the cosmological constant at this level.

Our calculation also means the following. We could have started with $R+R^2$ gravity right from the beginning, i.e., quadratic terms in curvature with independent coupling constants in addition to the Einstein-Hilbert action. Keeping the lowest order terms in $a$,
we get same action (at lowest order in $a$) for the larger cell scaled by only the volume under the renormalization group. This is like the case for a massless free theory driven by the Gaussian fixed point \cite{k}-\cite{m}, and providing the massless Gaussian critical surface. Our techniques  shift the focus to $R+R^2$ gravity as the massless Gaussian surface, from the non-interacting gravitons which are relevant only in the limit of vanishing gravitational coupling or infinitesimal curvature fluctuations. We are not bound by smallness of curvature fluctuations. The concern of renormalization group is to see how other coupling constants (including higher order contributions in $a$) flow away from this surface. This issue will be handled in a subsequent paper.

The techniques we have developed  have been used in the simplest way possible in this Paper. We can use them in more sophisticated ways. We can include  derivatives of curvature tensor in our choice of the metric Eq.\ref{mc}. We can specify the boundary conditions for each cell, and how they match with the those of the neighbors. We can match the coordinate charts of adjacent cells  in the process of coarse-graining. 

As we are working  with real-space techniques we  can take into account effects of exotic structures of Einstein gravity such as black holes, instantons and topologically non-trivial structures more easily. Matter fields also can be incorporated.

The problems with quantizing Einstein's theory has led to a wide variety of reactions, ranging from adding higher derivative terms to the action, or reformulating the theory (Loop Gravity), to adding new fields/symmetries (Supergravity/Superstrings).  Our techniques can be  useful  for these theories also, for calculating  behaviors at energies much lower than the Planck scale or symmetry breaking scales. 

\section{Appendix}\label{a}
Here we present calculation of the next order in $a$ contribution to the  Einstein-Hilbert  action for the cell. To be able to use manifest special relativity  symmetry, we do all calculations for the Euclidean case in a sphere of radius $a$ and take over the results for the Minkowski case.

We have,
\begin{eqnarray}
\sqrt{g}R=\sqrt{g}g^{ij}\Big{(}\frac{\partial}{\partial x^q}\Gamma^{q}\,_{ij}-\frac{\partial}{\partial x^j}\Gamma^{q}\,_{qi}
+\Gamma^{q}\,_{ij}\,\Gamma^{p}\,_{pq}
-\Gamma^{p}\,_{iq}\Gamma^{q}\,_{jp}\Big{)}.
\end{eqnarray}

We write the Taylor expansion formally,
\begin{eqnarray}
\Gamma^{q}_{ij}(x)=\Gamma^{q}_{ij}|_{0}+\Gamma^{q}_{ij}|_{1}+\Gamma^{q}_{ij}|_{2}+\Gamma^{q}_{ij}|_{3}+\cdots,
\end{eqnarray}
where the subscript denotes the order of powers of $x$ in the Taylor expansion. We have  \cite{n},
\begin{eqnarray}\label{g3}
&&\Gamma^{q}_{ij}|_{0}=0,
~~\Gamma^{q}_{ij}|_{1}=-\frac{1}{3}(R_{qijk}+R_{qjik})x^k,\nonumber\\
&&\Gamma^{q}_{ij}|_{2}=-\frac{1}{24}P_{ij}(5R_{qijk,l}+R_{qkil,j})x^kx^l\nonumber\\
&&\Gamma^{q}_{ij}|_{3}=P_{ij}\Big{(}(-\frac{1}{40}(3R_{qijk,lm}+R_{qkil,mj})\nonumber\\
&&+\frac{1}{360}(-23R_{qlnk}R_{nijm}+9R_{qinm}R_{nkjl}+15R_{qink}R_{nljm})\Big{)}x^k x^lx^m.
\end{eqnarray}
where $P_{ij}$ is a permutation in the indices.
As $\Gamma^{q}_{ij}|_{0}=0$ in normal coordinates,
the lowest order (in $a$) contribution from a 4-ball of radius $a$ to the action is,
\begin{eqnarray}
\sqrt{g}R|_0=\frac{\partial}{\partial x^q}\Gamma^{q}\,_{ii}|_1-\frac{\partial}{\partial x^i}\Gamma^{q}\,_{qi}|_1\nonumber\\
=-\frac{1}{3}(R_{qiiq}+R_{qiiq})+\frac{1}{3}(R_{qqii}+R_{qiqi}),
\end{eqnarray}
which using the properties of the Riemann tensor Eq.(\ref{rt1},\ref{rt2},\ref{rt3},\ref{rt4}) gives,
\begin{eqnarray}
\int_c \, d^4x \, \sqrt{g}R|_0=a^4 R.
\end{eqnarray}

For our choice of the metric fluctuations Eq.(\ref{m}), the $O(a^2)$ terms in the action are,
\begin{eqnarray}\label{a2}
\sqrt{g}R|_2=\frac{\partial}{\partial x^q}\Gamma^{q}\,_{ii}|_3-\frac{\partial}{\partial x^i}\Gamma^{q}\,_{qi}|_3
+\sqrt{g}|_2(\frac{\partial}{\partial x^q}\Gamma^{q}\,_{ii}|_1-\frac{\partial}{\partial x^i}\Gamma^{q}\,_{qi}|_1)\nonumber\\
+g^{ij}|_2(\frac{\partial}{\partial x^q}\Gamma^{q}\,_{ij}|_1-\frac{\partial}{\partial x^j}\Gamma^{q}\,_{qi}|_1)
+\Gamma^{q}\,_{ii}|_1\,\Gamma^{p}\,_{pq}|_1
-\Gamma^{p}\,_{iq}|_1\Gamma^{q}\,_{ip}|_1,
\end{eqnarray}
as $\sqrt{g}|_{1}$,$g^{ij}|_1$,$\Gamma^{q}\,_{ij}|_0$ are zero and $\Gamma^{q}\,_{ip}|_2$ does not contribute \cite{g}:
\begin{eqnarray}
&&\sqrt{g}|_{0}=1,~~\sqrt{g}|_{1}=0,
~~\sqrt{g}|_{2}=-\frac{1}{6}R_{ij}x^i x^j,
~~\sqrt{g}|_{3}=-\frac{1}{12}R_{ij,k}x^i x^jx^k\nonumber\\
&&\sqrt{g}|_{4}=\frac{1}{24}(-\frac{3}{5}R_{ij,k,l}+\frac{1}{3}R_{ij}
R_{kl}-\frac{2}{15}R_{imjn}R_{kmln})x^i x^jx^kx^l.
\end{eqnarray}
Also \cite{n},
\begin{eqnarray}
&&g^{ij}|_{0}=\delta_{ij},~~g^{ij}|_{1}=0,
~~g^{ij}|_{2}=\frac{1}{3}(R_{ikjl}+R_{iljk})x^kx^l,
~~g^{ij}|_{3}=\frac{1}{6}R_{ikjl,m}x^kx^lx^m\nonumber\\
&&g^{ij}|_{4}=(\frac{1}{20}R_{ikjl,mn}+\frac{1}{15}R_{rkil}R_{rmjn})x^kx^lx^mx^n.\\
\end{eqnarray}
Consider the terms quadratic in $\Gamma$ in Eq.\ref{a2},
\begin{eqnarray}
&&\Gamma^{q}\,_{ii}|_1\,\Gamma^{p}\,_{pq}|_1-\Gamma^{p}\,_{iq}|_1\Gamma^{q}\,_{ip}|_1\nonumber\\
&&=\frac{1}{9}((R_{qiik}+R_{qiik})(R_{ppql}+R_{pqpl})-(R_{piqk}+R_{pqik})(R_{qipl}+R_{qpil}))x^kx^l.
\end{eqnarray}
Now,
\begin{eqnarray}
\int_c d^4 x \, x^kx^l=v_2a^6\delta_{kl},
\end{eqnarray}
with
\begin{eqnarray}
v_2a^6=\frac{\pi^2}{4}\int_0^{b^2} dr^2 (r^2)^2=\frac{\pi^2}{12}b^6.
\end{eqnarray}
Therefore,
\begin{eqnarray}
\int_c d^4 x \, (\Gamma^{q}\,_{ii}|_1\,\Gamma^{p}\,_{pq}|_1-\Gamma^{p}\,_{iq}|_1\Gamma^{q}\,_{ip})|_1)=\frac{v_2a^6}{9}(R_{qiik}+R_{qiik})(R_{ppql}+R_{pqpk})=\frac{v_2a^6}{9}(\frac{5}{2}R_{ijkl}^2-R_{ij}^2).
\end{eqnarray}
As
\begin{eqnarray}
\frac{\partial}{\partial x^q}\Gamma^{q}\,_{ij}|_1-\frac{\partial}{\partial x^j}\Gamma^{q}\,_{qi}|_1
=\frac{2}{3}R_{ij},
\end{eqnarray}
\begin{eqnarray}
(\sqrt{g}g^{ij})|_2\big{(}\frac{\partial}{\partial x^q}\Gamma^{q}\,_{ij}|_1-\frac{\partial}{\partial x^j}\Gamma^{q}\,_{qi}|_1\big{)}
=\frac{v_2a^6}{9}(4R_{ij}^2-R^2)
\end{eqnarray}

Contribution to $\frac{\partial}{\partial x^q}\Gamma^{q}\,_{ii}|_3$ from various terms in Eq.\ref{g3} are as follows: 
\begin{eqnarray}
R_{qlnk}R_{nijm}x^k x^lx^m\rightarrow 4v_2a^6R_{ij}^2\nonumber\\
R_{qinm}R_{nkjl}x^k x^lx^m\rightarrow v_2a^6(-3R_{ijkl}^2-2R_{ij}^2)\nonumber\\
R_{qink}R_{nljm}x^k x^lx^m\rightarrow v_2a^6(-\frac{3}{2}R_{ijkl}^2+R_{ij}^2)
\end{eqnarray}
Contribution to $-\frac{\partial}{\partial x^i}\Gamma^{q}\,_{qi}|_3$ from various terms in Eq.\ref{g3} are as follows:
\begin{eqnarray}
R_{qlnk}R_{nijm}x^k x^lx^m\rightarrow v_2a^6(-\frac{3}{2}R_{ijkl}^2-R_{ij}^2)\nonumber\\
R_{qinm}R_{nkjl}x^k x^lx^m\rightarrow v_2a^6(\frac{3}{2}R_{ijkl}^2-R_{ij}^2)\nonumber\\
R_{qink}R_{nljm}x^k x^lx^m\rightarrow v_2a^6(\frac{3}{2}R_{ijkl}^2+R_{ij}^2)
\end{eqnarray}
Putting everything together,
\begin{eqnarray}
\int_c d^4x\,\sqrt{g}R|_2=v_2a^6\big{(}\frac{17}{160}R_{ijkl}^2+\frac{9}{40}R_{ij}^2-\frac{1}{9}R^2\big{)}
\end{eqnarray}


\begin{thebibliography}{1} \bibliographystyle{plain}

\bibitem{k} L.P. Kadanoff,  Scaling laws for Ising models near Tc, Physics {\bf 2} (1966) 263.
\bibitem{w} K. G. Wilson, Phys. Rev.  {\bf B4} (1971) 3174; 3184. 
\bibitem{kw} K. G. Wilson and I. G. Kogut, The renormalization group and the $\epsilon$-expansion, Phys. Rep.  {\bf 212} (1974) 75.
\bibitem{dg} {\it Phase Transitions and Critical Phenomena}, vol. 6, eds. C. Domb and M.S. Greene, Academic Press (1976).
\bibitem{m}  W.D. McComb, {\it Renormalization Methods: A Guide for Beginners}, Clarendon  Press, Oxford, 2004.

\bibitem{g} A. Gray, The volume of a small geodesic ball of a Riemannian manifold, Michigan.Math.J. {\bf 20}(1973) 329-344.
\bibitem{n} Wei-Tou Ni ,Geodesic Triangles and Expansion of Metrics in Normal Coordinates 
CHINESE JOURNAL OF PHYSICS {\bf 16} (1978) 223.

\bibitem{kpsk} A. Krasnitz, R. Potting, P. Sa´ and Y. A. Kubyshin (Eds.), {\it The Exact Renormalization Group, World Scientiﬁc, Singapore}, 1999.
\bibitem{p} Talks and Videos, {\it Renormalization Group Approaches to Quantum Gravity},  22-26 April 2014,  Perimeter Institute.
\bibitem{i} Slides, {\it 8th International Conference on the Exact Renormalization Group ERG2016 (smr 2826)}, 19-­23 September 2016, ICTP, Trieste.

\end{thebibliography}
\end{document}